\documentclass[global,twocolumn,final]{svjour}

\usepackage{latexsym}
\usepackage{epsfig}
\usepackage{rotating}
\usepackage{graphicx}
\usepackage{dcolumn}
\usepackage{bm}
\journalname{Applied Physics B}

\begin{document}

\title{Planar Ion Trap Geometry for Microfabrication}
\author{M. J. Madsen \inst{1} \and W. K. Hensinger \inst{1} \and D. Stick \inst{1} \and J. A. Rabchuk \inst{2} \and C. Monroe \inst{1}}
\institute{FOCUS Center and University of Michigan Department of Physics \and Western Illinois University} 

\date{4 January 2004}

\maketitle

\begin{abstract}
We describe a novel high aspect ratio radiofrequency linear ion trap geometry that is amenable to modern microfabrication techniques. The ion trap electrode structure consists of a pair of stacked conducting cantilevers resulting in confining fields that take the form of fringe fields from parallel plate capacitors. The confining potentials are modeled both analytically and numerically. This ion trap geometry may form the basis for large scale quantum computers or parallel quadrupole mass spectrometers. 

\end{abstract}

\section{\label{sec:intro}Introduction}

The ion trap has become an essential tool in several areas of physical science, including mass spectroscopy \cite{paul:1990}, atomic frequency standards \cite{fisk:1997}, precision atomic and molecular measurements \cite{vandyck:1987}, studies of fundamental quantum dynamics \cite{leibfried:2003} and quantum information science \cite{steane:1997,wineland:1998}. 
Many of these applications would benefit from miniaturized and multiplexed ion trap electrode structures well below the typical millimeter to centimeter scale.  Furthermore, smaller electrode dimensions offer the potential for stronger confining forces.

In this paper, we consider theoretically the electrical characteristics of a new type of micrometer-scale radiofrequency (RF) Paul ion trap fabricated using semiconductor micromaching and lithographic techniques such as micro-electro-mechanical-systems (MEMS) and molecular beam epitaxy (MBE). Such a device may enable new applications of ion trap technology such as ``quantum CCD'' scalable quantum computers \cite{kielpinski:2002}, optical cavity-QED with a localized single atom \cite{ye:1999,pinkse:2000,guthoehrlein:2002,mundt:2002}, and multiplexed quadrupole mass spectrometers that could be orders of magnitude smaller than previous devices \cite{taylor:2001}.

There has been much recent progress in the miniaturization of neutral atom electromagnetic trapping structures, involving, for example, micrometer-scale current-carrying wires on a substrate resulting in Bose-Einstein condensates on a microchip \cite{folman:2002}. Microscopic ion trap electrodes present their own challenges, as the confining forces are orders of magnitude stronger than those for neutral atom traps. Consequently, such ion traps will require greater control of unwanted or noisy electrode potentials, including the presence of thermal electric fields \cite{henkel:1999,turchette:2000}, residual charge on exposed insulating barriers, and ``patch'' potentials from inhomogeneities on the electrode surfaces \cite{wineland:1998,witteborn:1977}. None of these potential pitfalls appears fundamental, and such problems will only be overcome by testing various materials and approaches. We focus here on novel features of a proposed high aspect-ratio ion trap geometry and the resulting confining potentials.

The physical parameters of a model of the linear microtrap are discussed in Sec. \ref{sec:setup} along with a discussion of design considerations and issues with heating and power dissipation in semiconductor materials.  Section \ref{sec:CQP} contains a discussion of the RF ponderomotive potential of the linear microtrap model with results from numerical simulations of the potential. A geometrical efficiency factor is calculated, showing the performance of the linear microtrap as compared to an ideal quadrupole potential.  The static potential used for axial confinement in a linear trap is discussed in Sec. \ref{sec:asp} along with results from numerical simulations and comparison to an ideal hyperbolic trap.  The total potential along with examples of how to use the various geometric efficiency factors to calculate the trap frequencies of a given geometry are given in Sec. \ref{sec:total_pot}. The principal axes of the linear microtrap, which determine the efficiency of laser cooling ions in the linear microtrap, are evaluated in Sec. \ref{PA}. A method for rotating the axes for more efficient cooling is given.

\section{\label{sec:setup}Model Description}

\subsection{Basic Model}

\begin{figure}[ptbh]
\centering
\includegraphics[width=7cm,keepaspectratio]{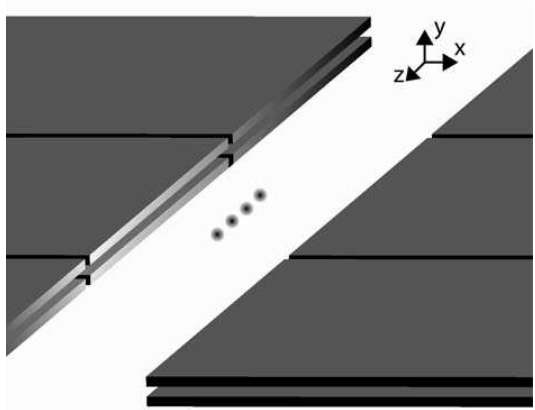}\caption{A three-dimensional drawing of the linear microtrap. A string of ions would lie along the $z$-axis as drawn.}
\label{fig:microtrap_autocad}
\end{figure}

The design of this new type of micrometer-scale RF trap is constrained by conventional semiconductor fabrication techniques, the need for clear laser optical access, and the characteristics of electrodes needed for linear traps. The design, illustrated in Fig. \ref{fig:microtrap_autocad}, is a two-layer planar geometry where both layers are divided into separate electrodes.  The division of each layer into six electrodes accommodates both the RF potentials and the static potentials needed to create a linear Paul trap \cite{raizen:1992}. This planar design is compatible with conventional photolithography techniques to define the electrode pattern. Each electrode is a cantilever anchored to an electrically isolated, conductive substrate and suspended from both sides of the planar structure. This ensures that there are no insulators near the center of the trap that could accumulate uncontrolled charge. Ions will be trapped in the space between the tips of each cantilever, along the $z$-axis in Fig. \ref{fig:microtrap_autocad}, near the center of the middle electrode.

\begin{figure}[ptbh]
\centering
\includegraphics[width=7cm,keepaspectratio]{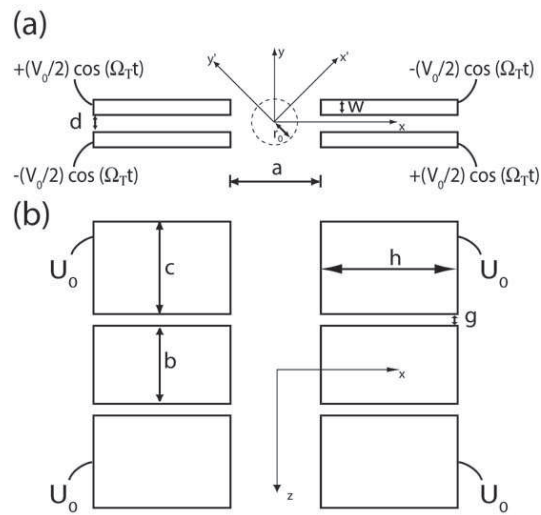}
\caption{(a) A schematic diagram of the linear microtrap design showing the side view.  The dimensions are labeled as are the RF voltages applied to the electrodes. (b) The top view of the linear microtrap with dimensions and static voltages as shown.}
\label{fig:microtrap_schematic}
\end{figure}

The cross-section of this linear microtrap (LMT) model at the center of the trap ($z=0$) is shown in Fig. \ref{fig:microtrap_schematic}(a). The thickness of each layer is labeled $w$; the layer separation is $d$; the tip-to-tip separation of the cantilevers is $a$.  Two ratios are useful for characterizing the behavior of the electric potentials: the trap aspect ratio, or the ratio of the tip-to-tip cantilever separation to the layer separation $\alpha=a/d$, and the ratio of the layer separation to the layer thickness $\delta=d/w$.  An RF voltage is applied between each set of diagonally opposing electrodes as shown in Fig. \ref{fig:microtrap_schematic}(a).

A top view of the linear microtrap model is shown in Fig. \ref{fig:microtrap_schematic}(b). The width of the center cantilevers along the $z$-axis of the trap is labeled $b$; the width of the end-cap cantilevers is $c$; and the length of the cantilevers in the model is $h$. In order to electrically insulate the center from the end-cap cantilevers, a small gap is introduced of width $g$. This allows for separate potentials to be applied to all twelve cantilevers, or electrodes.  Static voltages are applied to both layers on the four end-cap electrodes on either side of the center cantilevers to provide axial confinement, as shown in Fig. \ref{fig:microtrap_schematic}(b).

The potentials of the LMT can be separated into two parts for analysis. The first part is the ponderomotive potential generated by the RF voltages.  In the limit where gap width $g$ is much smaller than $a$, $b$ and $c$ (Fig \ref{fig:microtrap_schematic}(b)), the RF potential is approximately independent of $z$ near the center of the trap. In the cross-sectional plane at $z=0$, this RF potential generates a two-dimensional trapping pseudopotential and is discussed in Section \ref{sec:CQP}. The second part is the potential generated by applying static voltages to the end-cap  electrodes.  This potential provides axial confinement for ions in the center of the trap and is described in Section \ref{sec:asp}. Note that the end-cap electrodes have both the RF voltages applied to reduce the $z$ dependence of the RF field near the center of the trap and static voltages to create the end-caps. The center electrodes are all assumed to be held at static ground.

\subsection{Fabrication Considerations}

The linear microtrap model is designed to simulate a trap design that can be fabricated using conventional micro-processing techniques. The sizes of the electrode features that will be analyzed in this model are typical of current fabrication processes. There are several different processes that could be used to fabricate these microtraps: a) Silicon-based microelectromechanical machining (MEMS) techniques; b) Gallium-Arsenide (or other suitable material) based molecular-beam epitaxy (MBE) grown wafers and associated etching processes; or c) other relevant techniques such as anodic wafer bonding or flip-chip technologies. The length of the cantilevers is limited by allowable mechanical vibrations in the cantilevers themselves, as well as limits to the mechanical stability of the cantilevers under electromechanical forces due to the applied RF and static voltages. The mechanical forces exerted on the cantilevers can be approximated using structural cantilever analysis \cite{cleveland:1993}. Following this analysis, the spring constant of the center rectangular cantilever can, for example, be expressed as
\begin{equation}
k=E\left(\frac{w^{3}b}{4h^{3}}\right)
\label{eqn:k1}
\end{equation}
where $E$ is the Young's Modulus of the relevant material. The force on one cantilever due to an applied potential difference $V_{0}$ between layers can be approximated as the gradient of the potential in a parallel plate capacitor of area $A=hb$ and plate separation $d$.
\begin{eqnarray}
F&=&-\frac{\partial U_{Capacitor}}{\partial d}\nonumber\\
&=&-\frac{\epsilon_{0}}{2}\frac{\partial}{\partial d}\left(  \frac{hbV_{0}^{2}}{d}\right)\nonumber\\
&=&\frac{\epsilon_{0}}{2}\left( \frac{hbV_{0}^{2}}{d^{2}}\right)
\label{eqn:force1}
\end{eqnarray}
Although the actual force is distributed across the length of the capacitor, by approximating the force as being concentrated at the tip, one can find an upper bound on the cantilever tip deflection.  Treating the cantilever as a classical spring with the force applied at the tip, and using the spring constant from Eq. \ref{eqn:k1}, the maximum tip deflection $x_d^{(0)}$ can be approximated as
\begin{equation}
x_{d}^{(0)}\sim\frac{2\epsilon_{0}h^{4}V_{0}^{2}}{Ed^{2}w^{3}}.
\label{eqn:xdef1}
\end{equation}
A typical deflection for a GaAs cantilever with $E=85.5$GPa and dimensions $h=100\mu$m, $d=2\mu$m, $w=2\mu$m, with an applied voltage difference of $V_{0}=20$V, is $x_{d}^{(0)}=260$nm. The resonant frequency of the cantilever can also be calculated \cite{cleveland:1993} as a function of the material density $\rho$, the Young's Modulus, the cantilever width $w$ and the length $h$:
\begin{equation}
\omega_{vib}/2\pi=0.162\sqrt{E/\rho}\frac{w}{h^{2}}\label{eqn:cantfreq}
\end{equation}
which for GaAs ($\rho=5.31$gm/cm$^{3}$) is $\omega_{vib}/2\pi\approx130$kHz for the same dimensions as previously discussed. 

For an RF potential $V_0\cos(\Omega_{T}t)$ applied to the cantilever electrodes, the amplitude of the tip deflection in Eq. \ref{eqn:xdef1} is expected to be further reduced by a Lorentzian factor of $\omega_{vib}^2/\Omega_{T}^2\ll1$. Here, it is assumed that the RF frequency is far from resonance, or $\Omega_{T}\gg\omega_{vib}/Q$, where $Q\gg1$ is the quality factor of the mechanical resonance \cite{lifshitz:2000}. While the above electromechanical forces do not appear troublesome, the actual forces may be considerably higher due to free charges on the electrode layers that are driven by the applied potentials.  In any case, it may be necessary to isolate the cantilevered electrodes from noisy electrical signals near the mechanical resonance.

The trap strength may be limited by the maximum voltage that can be applied to the electrodes before the occurrence of electric field break-down. The theoretical limit to the breakdown voltage is dependent on the bandgap of the semiconductor material and, for Si and GaAs, is on the order of 40-50 V/$\mu$m \cite{david:1996} and for silicon nitride, on the order of 300 V/$\mu$m \cite{rauthan:1992}. For a layer separation of 2$\mu m$, the maximum applied voltage is expected to be of order $V_{0}=100$V.

\subsection{RF Dissipation and Thermal Fields}

The fabrication considerations for the implementation of this new type of linear microtrap suggest that highly doped semiconductors could be used as electrodes. Because doped semiconductors have a resistivity several orders of magnitude greater than the metal conductors typically used in ion traps, it is necessary to estimate the power dissipation of the microtrap due to RF losses in the cantilevers.  Additionally, the finite conductivity of semiconductor materials will lead to thermal electric fields that will generate heating of the quantized motion of ions in the center of the trap.  

The RF dissipation can be estimated with a simple model of lumped circuit elements, since the trap structure is much smaller then the RF wavelength. Each RF electrode is modeled as a small series resistance $R$ shunted by a capacitance $C$ at the trap; inductance of the electrodes is assumed negligible compared to $1/(C\Omega_{T}^{2})$.  In addition, RF loss in the insulator separating the electrodes contributes to a parallel resistance characterized by the loss tangent $\tan\delta$.  Assuming $RC\Omega_T, \tan\delta \ll 1$, the power loss is
\begin{equation}
P_{d}=\frac{V_0^2 C \Omega_T}{2}(R C \Omega_T + \tan\delta).
\label{eqn:power_disp}
\end{equation}
For values envisioned here, $V_{0} \sim 20$V at $\Omega_{T}/2\pi \sim 50$MHz, $C \sim 10$pF, $\tan\delta \sim 0.0002$ and $R \sim 10\Omega$, resulting in a power dissipation of $P_{d} \sim 40$mW per electrode.

Additionally, Johnson noise in the electrodes will generate thermal electric fields that will cause heating of the quantized ion motion.  A simple model can be used to calculate the heating due to the resistivity of the trap electrodes \cite{wineland:1998,turchette:2000}. For an ion held at a distance $z$ from a conductive plane, the heating rate is given by
\begin{eqnarray}
\frac{\partial{E}}{\partial t}&=&\hbar\omega\dot{\bar{n}}\nonumber\\
&=&\frac{e^2k_BTR(\omega_{s})}{mz^2}
\label{eqn:heating_quanta}
\end{eqnarray}
where $\omega$ is the secular frequency and $\bar{n}$ is the average vibrational quantum number of an ion in the trap. In the limit where the conductor thickness $w$ is much smaller than the distance to the ion $z$, and both dimensions are smaller then the skin depth $\delta$ of the conductor ($w\ll z\ll\delta$), the resistance $R$ in Eq. \ref{eqn:heating_quanta} is frequency independent: $R\approx\rho z/(zw)$, where $\rho$ is the material resistivity.  Here, the effective volume of the conductor contributing to the thermal fields is of order $z^2w$. Again, using typical values for doped semiconductors, the skin depth $\delta$ is a few hundred micrometers, the thickness of the conductor is 2$\mu$m and the ion is $20\mu$m from the conductor.  In this limit, using a secular frequency of $\omega_{s}/2\pi=10$MHz, and ${}^{111}$Cd${}^{+}$ ions, Eq. \ref{eqn:heating_quanta} predicts a thermal heating rate of about 10 quanta/sec. Since this model pertains to fluctuating uniform thermal electric fields from a single conducting plane, the actual thermal electric fields are expected to be much smaller because the trap structure surrounds the ion with a high degree of symmetry, resulting in some degree of cancellation of thermal fields from opposite electrodes.  In any case, the heating rate will likely be limited in practice by fluctuating patch fields on the electrode surfaces \cite{turchette:2000}.

\section{\label{sec:CQP} RF Ponderomotive Potentials}

\subsection{Time-dependent RF potentials}
As described above, the analysis of the potentials in a linear RF Paul trap can be divided into the transverse RF trap generated by RF voltages applied to the appropriate electrodes, and the axial trap and transverse anti-trap generated by static voltages applied to the end-cap electrodes.  Focusing first on the time-varying potential generated by the RF voltages, the analysis can be simplified by using a pseudopotential approximation. The motion of an ion in an RF potential of the form
\begin{equation}
\Phi(x,y,z,t)=V(x,y,z)\cos(\Omega_{T}t)
\end{equation}
can be approximated using a ponderomotive pseudopotential \cite{dehmelt:1967}:
\begin{equation}
\psi=\frac{e^{2}}{4m\Omega_{T}^{2}}\left|  \nabla V(x,y,z)\right|^{2}
\label{eqn:ponder_defn}
\end{equation}
Ion motion is in the pseudopotential can be approximated as secular harmonic motion \cite{wineland:1998} with frequency
\begin{equation}
\omega_{p}^{2}=\frac{e^{2}}{4m^{2}\Omega_{T}^{2}}\frac{\partial^{2}%
}{\partial{x^{2}}}\left(  \left|  \nabla V(x,y,z)\right|  ^{2}\right).
\end{equation}
The micromotion due to the time dependence of the RF potential is small in the limit where $q\equiv 2\sqrt{2} \omega_{p}/ \Omega_{T} \ll 1$ \cite{dehmelt:1967}.

Since the secular ion motion is dependent only on the gradient of $V(x,y,z)$, it is possible to calculate the effective (or ponderomotive) potential of the linear microtrap using an electrostatic analysis. Moreover, since the RF potential is approximately uniform along the $z$-axis near the center of the trap, it can be described in the $z=0$ plane as a function only of $x$ and $y$, reducing the calculation of the RF potential to two dimensions.

\subsection{\label{sec:two-d-hyp}Hyperbolic Electrode Model}

\begin{figure}[ptbh]
\centering
\includegraphics[width=7cm,keepaspectratio]{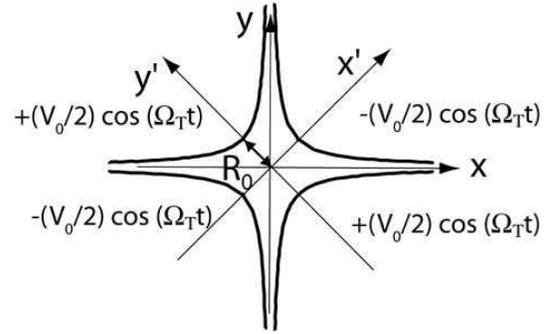}
\caption{The hyperbolic electrode geometry is used as a basis for comparing the linear microtrap.  The characteristic dimension of the hyperbolic electrode geometry is the radius $R_{0}$ as shown.}
\label{fig:two_d_hyperbolic}
\end{figure}
One common configuration of a linear Paul trap consists of four infinitely long hyperbolic electrodes. This hyperbolic electrode model will be used as a standard of comparison for the linear microtrap. The cross-section of hyperbolic electrodes with a characteristic radius $R_{0}$ is shown in Fig. \ref{fig:two_d_hyperbolic}. For the potentials applied according to Fig. \ref{fig:two_d_hyperbolic}, the exact potential amplitude is
\begin{eqnarray}
V_{\mathrm{hyp}}(x^{\prime},y^{\prime}) &= &\frac{V_{0}}{2R_{0}^{2}}\left(
x^{\prime}{}^{2}-y^{\prime}{}^{2}\right)\\
&=&\frac{V_{0}}{2}\frac{r^2}{R_{0}^{2}}\cos2\theta^{\prime}\nonumber.
\label{eqn:cross_hyp_pot}
\end{eqnarray}
where the coordinate system $(x^{\prime},y^{\prime})$ is indicated in Fig. \ref{fig:two_d_hyperbolic}. 

The pseudopotential that corresponds to this hyperbolic potential is calculated using Eq. \ref{eqn:ponder_defn}.
\begin{equation}
\psi_{\mathrm{hyp}}=\frac{e^{2}V_{0}^{2}}{4m\Omega_{t}^{2}R_{0}^{4}}\left(
x^{\prime}{}^{2}+y^{\prime}{}^{2}\right)
\label{eqn:cross_hyp_ponder}
\end{equation}
The secular frequency of a ion moving in this ponderomotive pseudopotential is therefore
\begin{eqnarray}
\omega_{p,\mathrm{hyp}}&=&\frac{eV_{0}}{\sqrt{2}m\Omega_{t}R_{0}^{2}}\nonumber\\
&=&\sqrt{\frac{eV_0q}{4mR_0^2}}.
\label{eqn:cross_hyp_freq}
\end{eqnarray}

\subsection{Linear Microtrap Transverse Potential Analysis}

The microtrap potential amplitude $V_{_{\mathrm{LMT}}}$ is computed near the center of the trap. This potential is then decomposed as an infinite set of cylindrical harmonics \cite{syms:1998}:
\begin{eqnarray}
V_{_{\mathrm{LMT}}}(r,\theta^{\prime})&=&V_0\left[\sum\limits_{m=1}^{\infty}
{C_{m}\left(  r/r_{0}\right)  ^{m}\cos(m\theta^{\prime})}\right.\nonumber\\
&&\left.+\sum\limits_{n=1}
^{\infty}{S_{n}\left(  r/r_{0}\right)  ^{n}\sin(n\theta^{\prime})}\right]
\label{eqn:phi_decomp}
\end{eqnarray}
where $C_{m}$ and $S_{n}$ are expansion coefficients and $\theta^{\prime}$ is taken as the angle from the $x^{\prime}$ axis. The characteristic radius over which the potential is approximated by this expansion is $r_{0}$.

The $C_2$ coefficient provides a comparison between the potential of the linear microtrap and the quadrupole potential of the hyperbolic electrode geometry of radius $r_0$.  Other nonzero coefficients in the expansion of Eq. \ref{eqn:phi_decomp} describe the anharmonic character of the microtrap potential. Symmetry considerations reduce the number of terms allowed in the expansion. Given the potential amplitude of $\pm V_{0}/2$ applied to opposite electrodes as shown in Fig. \ref{fig:microtrap_schematic}(a), the potential is antisymmetric along the lines $x=0$ and $y=0$ and symmetric in reflection about the origin leading to the only non-zero terms in Eq. \ref{eqn:phi_decomp} as $m=2,6,10,\ldots$ and $n=4,8,12,\ldots$.

The expansion coefficients are calculated by numerically evaluating the LMT potential using finite element analysis or other appropriate numerical field simulators and calculating the overlap integrals within a circle of radius $r_0$ of the potential $V_{_{\mathrm{LMT}}}$ with the cylindrical harmonics $\left( r/r_{0}\right) ^{m} \cos(m\theta^{\prime})$ and $\left(r/r_{0}\right) ^{n} \sin(n\theta^{\prime})$ \cite{syms:1998}.

A geometric efficiency factor $\eta$ can be used to compare the microtrap potential with the quadrupole potential of the hyperbolic electrodes of comparable size. The size of the linear microtrap is given by the distance from the center of the trap to the nearest point on the tip of the electrodes $\ell_{\mathrm{eff}}\equiv\sqrt{(a/2)^{2}+(d/2)^{2}}$. Then, $\eta$ is defined as the ratio of the quadrupole part of the potential generated by the LMT $V_{_{\mathrm{LMT}}}^{(2)}$ and a hyperbolic trap with $R_{0}=\ell_{\mathrm{eff}}$.

\begin{equation}
\eta=\frac{ V_{_{\mathrm{LMT}}}^{(2)} }{V_{\mathrm{hyp}}}=\frac{{2C_{2}\ell_{\mathrm{eff}}^{2}} 
}{{r_{0}^{2}}}.
\label{eqn:eta_LMT}
\end{equation}
The quadrupole portion of the linear microtrap can therefore be written in a form differing from the hyperbolic electrode potential (Eq. \ref{eqn:cross_hyp_pot}) by only the geometric factor $\eta$.
\begin{equation}
V_{_{\mathrm{LMT}}}^{(2)}(x',y')=\frac{V_{0}\eta}{2\ell_{\mathrm{eff}}^{2}} \left(  x'{}^{2}-y'{}^{2}\right) 
\label{eqn:cross_LMT_pot}
\end{equation}
The ponderomotive potential for the microtrap can then be evaluated using Eq. \ref{eqn:ponder_defn}: 
\begin{equation}
\psi_{_{\mathrm{LMT}}}=\frac{e^{2}V_{0}^{2}\eta^{2}}{4m\Omega_{T}^{2}\ell_{\mathrm{eff}}^{4}}\left(  x^{2}+y^{2}\right).
\label{eqn:cross_LMT_ponder}
\end{equation}
Finally, the effective secular frequency of an ion in the linear microtrap is only modified by the factor $\eta$ from the form of the secular frequency in the trap due to the hyperbolic electrodes (Eq. \ref{eqn:cross_hyp_freq}). With this form of the secular frequency, one can compare the trap strength and performance of the linear microtrap.
\begin{equation}
\omega_{p_{,\mathrm{LMT}}}=\frac{eV_{0}\eta}{\sqrt{2}m\Omega_{T}\ell_{\mathrm{eff}}^{2}}
\label{eqn:cross_LMT_freq}
\end{equation}

\begin{figure}[ptbh]
\centering
\includegraphics[width=7cm,keepaspectratio]{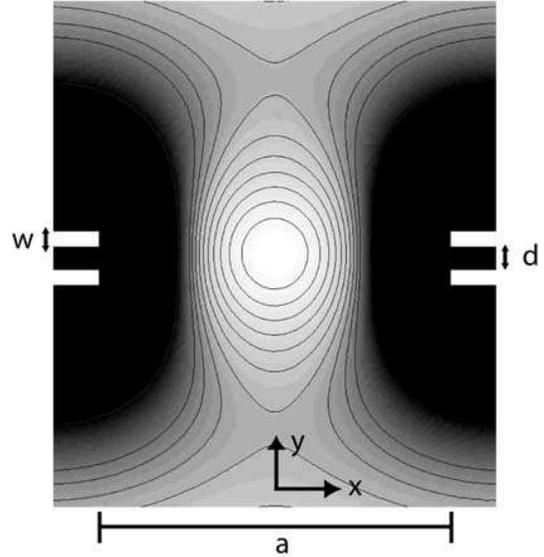}
\caption{Equipotential lines of the pseudopotential $\psi_{_{\mathrm{LMT}}} $ in the
$z=0$ plane for aspect ratio $\alpha=10$ and ratio of layer separation to layer thickness of $\delta=1$. The ponderomotive potential reaches a maximum along the $y$ axis near $\ell_{\mathrm{eff}}$. The contour lines are spaced on a linear scale and are shown to illustrate the circular nature of the ponderomotive potential at the center of the trap.  The gray-scale shading is also on a linear scale.}
\label{fig:pseudo_pot}
\end{figure}

The equipotential lines of the calculated ponderomotive potential are shown in Fig. \ref{fig:pseudo_pot} along with the potential magnitude indicated by a gray-scale. Note that, although the cantilever geometry does not have cylindrical symmetry, the pseudopotential is approximately circular within a distance on the order of one-eighth the tip-to-tip separation $a$ as will be shown from the numerical results in Sec. \ref{sec:FEAM} where $C_2$ is found to be the dominant term in the expansion at this distance from the center.

\subsection{\label{sec:FEAM}Finite Element Analysis Method}
\begin{figure}[ptbh]
\centering
\includegraphics[width=7cm,keepaspectratio]{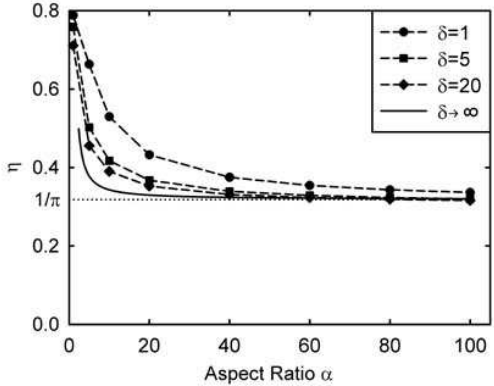}
\caption{The ponderomotive potential geometric efficiency factor $\eta$ as a function of the ratio of the tip-to-tip separation to the layer separation: the aspect ratio $\alpha$. The other degree of freedom is the ratio of the electrode separation to the layer thickness, $\delta=d/w$.  The solid line is an analytic solution for $\eta$ found using complex analysis techniques with $\delta\rightarrow\infty$ and is valid for $\alpha \gg 1$}
\label{fig:eta_alpha}
\end{figure}

The class of finite element analysis solvers that is used here divides a two-dimensional space into a series of triangles to calculate the linear microtrap potential. The two-dimensional finite element analysis package in Matlab version 6.5 was used to calculate the RF potentials. The results were compared with the two-dimensional projection of potentials calculated using two different three-dimensional finite element analysis packages, Maxwell 3D from Ansoft, and Opera 3D from VectorFields, and found consistent. The field is approximated at each vertex on the triangles, then an interpolation is made within each element to calculate the field on an rectangular grid. Different trap configurations are analyzed using the method described above and the ratio $\eta$ of the microtrap potential to the quadrupole hyperbolic potential is shown in Fig. \ref{fig:eta_alpha} evaluated at a radius of $r_0=a/8$. The uncertainty of the simulation data is less then 5\% and is due primarily to a finite grid spacing and the finite bounding box size. The solid line in the figure is an analytic solution for cantilevers of infinitesimal thickness.

One can see that as the trap aspect ratio $\alpha=a/d$ increases, the geometric factor $\eta$ approaches a constant, non-zero value.  The asymptotic value can be evaluated using complex analysis techniques and is described in Appendix \ref{sec:analytic}. The result from Eq. \ref{eqn:analytic_eta} for large $\alpha$ is $\eta=1/\pi$. Additionally, as the aspect ratio approaches one, the trap becomes more like the hyperbolic electrode geometry.  The other degree of freedom of the linear microtrap is the ratio of the layer separation to the layer thickness, $\delta=d/w$.  Note that the strength of the microtrap decreases as the layer thickness decreases with respect to the layer separation.

\begin{figure}[ptbh]
\centering
\includegraphics[width=7cm,keepaspectratio]{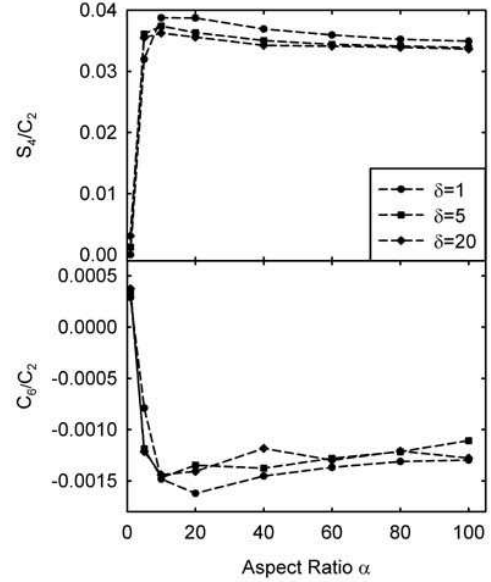}
\caption{The two largest higher-order terms of the expansion in Eq. \ref{eqn:phi_decomp} shown as a ratio over $C_2$ for various trap aspect ratios $\alpha=a/d$ and given as a function of the layer separation over the layer thickness $\delta=d/w$ evaluated at $r_0=a/8$}
\label{fig:higher_coeff_r0}
\end{figure}

The higher-order coefficients of the expansion shown in Eq. \ref{eqn:phi_decomp} for the potential $V_{_{\mathrm{LMT}}}$ are shown in Fig. \ref{fig:higher_coeff_r0}.  The dominant higher-order term is $S_4$, which, at a fixed radius of $r_0=a/8$, is only a few percent of $C_2$.  The two next largest terms are also shown although the magnitude is small enough to be negligible when considering ion motion. The relationship between the $C_2$ and the next three largest terms of the expansion as a function of the aspect ratio $\alpha$ and $\delta$ is shown in Fig \ref{fig:higher_coeff_r0}.  Coefficients $S_4$, $C_6$ appear to approach an asymptotic value as the trap aspect ratio increases. The ratios of all higher-order terms to the coefficient $C_2$ ($C_m/C_2$ and $S_n/C_2$) for $m,n>6$ are less than $10^{-3}$.

\begin{figure}[ptbh]
\centering
\includegraphics[width=7cm,keepaspectratio]{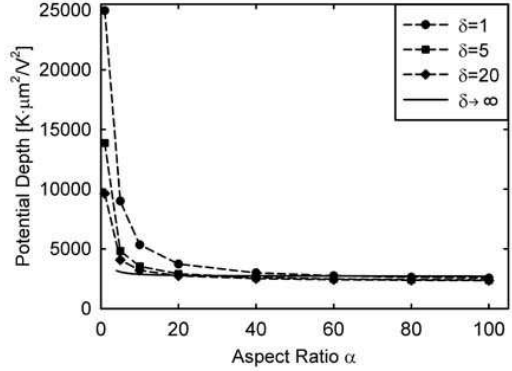}
\caption{The scaled trap depth as a function of the trap aspect ratio $\alpha=a/d$ and the ratio of layer separation to layer thickness $\delta=d/w$.  The trap depth is scaled to the tip-to-tip separation $a$, in micrometers and to the applied voltage $V_0$. The analytic result is shown as a solid line with $\delta\rightarrow\infty$ and is valid for $\alpha\gg1$.}
\label{fig:trap_depth}
\end{figure}

The absolute depth of the ponderomotive RF trap is also of interest when considering ion loading and collisions with background gas.  The trap depth is defined as the maximum height of the ponderomotive potential barrier along the weak axis of the trap and is plotted in Fig. \ref{fig:trap_depth}.  A trap frequency of $\Omega_{T}/2\pi=50$MHz and the mass of $^{111}$Cd$^{+}$ were used to calculate the depth, given in scaled units of [K$\cdot\mu$m$^2$/V$^2$]. To find the depth of a specific trap, the data must be multiplied by the applied voltage $V_0^2$ in [V$^2$] and divided by the square of the absolute tip-to-tip separation $a^2$ in [$\mu$m$^2$]. The depth asymptotically approaches a constant value of approximately $2400$K$\cdot\mu$m$^2$/V$^2$ for large cross-sectional aspect ratio as can be found from the analytic solution (Eq. \ref{eqn:analytic_depth}). The size of the ponderomotive trap $r_{\mathrm{max}}$ is characterized by either the distance of the maximum in the ponderomotive potential from the center of the trap or $a/2$, whichever is smaller. As the trap aspect ratio increases $r_{\mathrm{max}}$ is determined by the maximum in the RF pseudopotential along the $y$-axis and is approximately half the tip-to-tip electrode separation $0.5a$. 

\begin{figure}[ptbh]
\centering
\includegraphics[width=7cm,keepaspectratio]{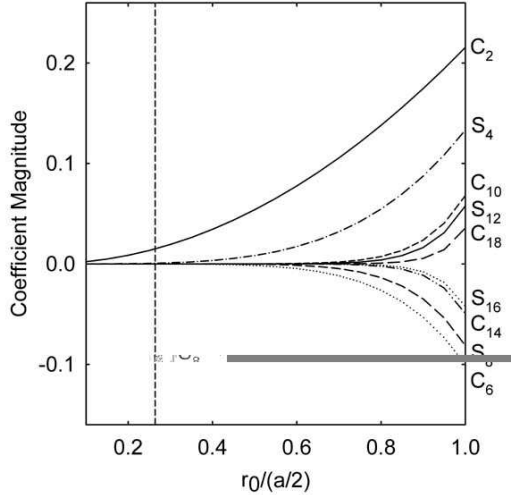}
\caption{The dependence on the expansion coefficients $C_m$ and $S_n$ as a function of $r_0/(a/2)$.  The higher order terms become significant as the overlap integrals cover more of the area between the electrodes. The geometry used was $\alpha=20$ and $\delta=1$, a worst case scenario from Fig. \ref{fig:higher_coeff_rmax}.  The dashed vertical bar indicates $r_0=a/8$.}
\label{fig:coeff_r0_dep}
\end{figure}

\begin{figure}[ptbh]
\centering
\includegraphics[width=7cm,keepaspectratio]{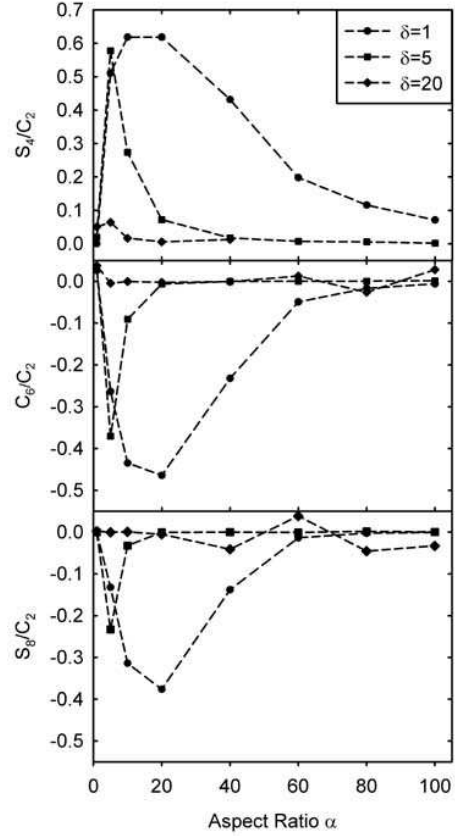}
\caption{The three largest higher-order terms of the expansion in Eq. \ref{eqn:phi_decomp} evaluated within a radius $r_0=r_{\mathrm{max}}$, where $r_{\mathrm{max}}$ is the maximum of the ponderomotive potential. The coefficients are shown as a percentage of the largest term $C_2$ for various trap aspect ratios $\alpha=a/d$ and given as a function of the layer separation over the layer thickness $\delta=d/w$ with an error of 5\%.}
\label{fig:higher_coeff_rmax}
\end{figure}

Since the ponderomotive potential within the region $r<r_{\mathrm{max}}$ will trap ions, the expansion of the potential from Eq. \ref{eqn:phi_decomp} within that entire area is also of interest. The expansion of the potential within a circle of radius $r_0=r_{\mathrm{max}}$ contains a larger contribution from the higher-order coefficients than an expansion fixed at $r_0=a/8$ as illustrated in Fig. \ref{fig:coeff_r0_dep}. The higher-order coefficients for the expansion of the linear microtrap potential are shown in Fig. \ref{fig:higher_coeff_rmax}, evaluated at $r_0=r_{\mathrm{max}}$.

\subsection{Residual Axial Ponderomotive Potential}

The previous analysis is based on the assumption that the linear microtrap electrodes are infinitely long in the $z$-dimension.  However, the actual trap has finite electrode lengths, labeled $b$ and $c$ in Fig. \ref{fig:microtrap_schematic}(b), which together with the small electrode gaps (labeled $g$ in Fig. \ref{fig:microtrap_schematic}) lead to a small ponderomotive potential in the $z$ direction. The magnitude of this axial ponderomotive potential can be compared to the transverse ponderomotive potential $\psi_{_{\mathrm{LMT}}}$ of Eq. \ref{eqn:cross_LMT_ponder}. To find the axial contribution, the entire three-dimensional RF potential $V_{_{\mathrm{LMT}}}(x,y,z)$ must be computed. Once found, one can use the ponderomotive potential approximation Eq. \ref{eqn:ponder_defn} to calculate the trap frequency along the $z$-axis.

The gradient of the three-dimensional potential is found, then the pseudopotential is evaluated.  A Taylor expansion of the pseudopotential along the $z$ axis (about $z=0$) gives the coefficient for the harmonic $z^{2}$ term in the ponderomotive potential:
\begin{equation}
H_{z}=\frac{1}{2}\frac{\partial^{2}}{\partial z^{2}}\left(  \left|  \nabla
V_{_{\mathrm{LMT}}}(x,y,z)\right|  ^{2}\right).
\end{equation}
The details of the three-dimensional potential calculation are given below, but the method is similar to the two-dimensional finite difference analysis. Typically the data is extracted along the $z$ axis and then fit to a quadratic polynomial to find the coefficient $H_z$. This coefficient allows one to make a comparison between the quadrupole trapping pseudopotential in the $z=0$ cross-sectional plane, and the ponderomotive potential along the $z$-axis.  This three-dimensional ponderomotive potential is similar to the transverse potential of Eq. \ref{eqn:cross_LMT_ponder} with the addition of the $z^2$ term.
\begin{equation}
\psi_{_{\mathrm{LMT}}}(x,y,z)=\frac{e^{2}V_{0}^{2}\eta^{2}}{4m\Omega_{T}^{2}\ell_{\mathrm{eff}}^{4}
}\left(  x^{2}+y^{2}+\sigma_{z}z^{2}\right),
\end{equation}
where $\sigma_{z}=H_{z}\ell_{\mathrm{eff}}^{4}/\eta^{2}$ is the ratio of the residual axial to transverse ponderomotive potential. The resulting frequency along the $z$-axis is $\omega_{z}=\sqrt{\sigma_{z}}\omega_{p_{,\mathrm{LMT}}}$.

\begin{figure}[ptbh]
\centering
\includegraphics[width=7cm,keepaspectratio]{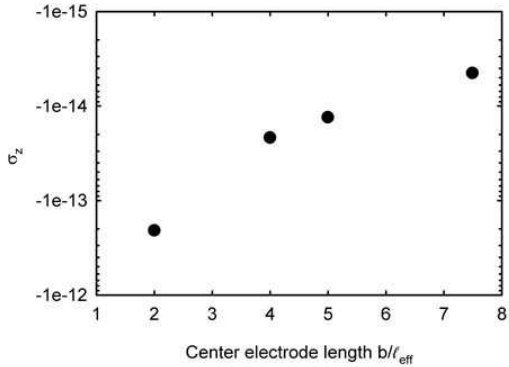}
\caption{The ratio of the residual axial frequency to the transverse ponderomotive frequency $\sigma_{z}$ as a function of the center electrode length.  The end-cap electrodes were fixed at $5\ell_{\mathrm{eff}}$ with a fixed gap spacing of $1/10\ell_{\mathrm{eff}}$, $\alpha=20$, and $\delta=1$.}
\label{fig:sigmaz_beta}
\end{figure}

\begin{figure}[ptbh]
\centering
\includegraphics[width=7cm,keepaspectratio]{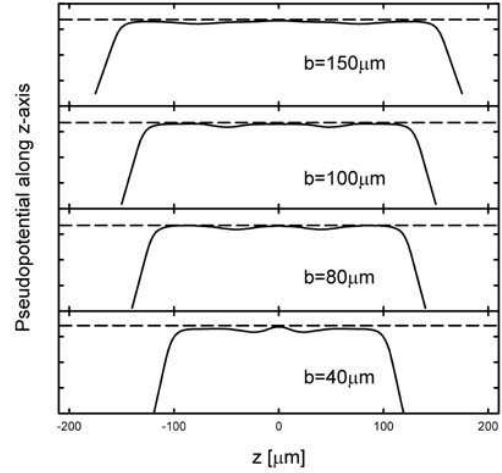}
\caption{Illustration of the change in the residual axial ponderomotive potential for various center electrode lengths ($b$).  The potential along the $z$-axis is shown for various center electrode lengths where the end-cap electrodes have been fixed at $100\mu$m.}
\label{fig:psi_z_axis}
\end{figure}

The results from the numerical simulation in Fig. \ref{fig:sigmaz_beta} are given for a cross-sectional aspect ratio of $\alpha=20$ and for $\delta=1$ (ratio of the layer separation to the layer width). The ponderomotive potential along the $z$-axis is shown in Fig. \ref{fig:psi_z_axis} to illustrate the degree to which the notch gap $g$ contributes to the residual potential at the center of the trap. Since $\sigma_{z}\ll1$, the ponderomotive contribution to the potential along the $z$-axis can be neglected.

\section{\label{sec:asp}Static Potentials}

\subsection{Hyperbolic Geometry}

\begin{figure}[ptbh]
\centering
\includegraphics[width=7cm,keepaspectratio]{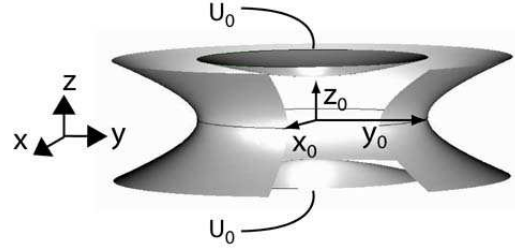}%
\caption{Three-dimensional hyperbolic electrodes are shown here. The electrodes along the $z$-axis are held at a voltage of $U_0$, while the center electrode is grounded. The potential has an elliptical cross-section in the $xy$ plane corresponding to $\epsilon=0.86$ and, for $U_0>0$, is trapping along the $z$-axis, but anti-trapping along $x$ and $y$, valid for $0<\epsilon<1$.}
\label{fig:three_d_hyperbolic}
\end{figure}

Like the two-dimensional potential in Sec. \ref{sec:two-d-hyp}, the static potential used to confine the ions along the $z$-axis in the linear microtrap can be compared to a three-dimensional idealized hyperbolic electrode potential.  Figure \ref{fig:three_d_hyperbolic} shows an elliptical hyperbolic electrode geometry where $x_0$, $y_0$, and $z_0$ are the distances along the principal axes of the ellipse from the center of the trap to the electrodes. The potential within the electrodes, up to a constant term, is 
\begin{equation}
U_{\mathrm{hyp}}=\frac{U_{0}}{s^{2}}\left(  -\epsilon x^{2}-(1-\epsilon)y^{2}+z^{2}\right)
\label{eqn:axial_hyp_pot}
\end{equation}
where $s^2=z_0^2+\epsilon x_0^2$ and $\epsilon x_0^2=(1-\epsilon)y_0^2$. The geometric anisotropy factor $\epsilon$ is related to the eccentricity of various conic sections that can be superimposed on the three-dimensional hyperbolic electrode structure.  The special case where $\epsilon=1/2$ corresponds to circular symmetry in the $xy$ plane. For values of $0<\epsilon<1$ and $U_0>0$, the potential is trapping in $z$ and anti-trapping in the $xy$ plane, as shown in the figure for $\epsilon=0.86$.  Outside of that range, the axes in the figure must be rotated to describe the potential of Eq. \ref{eqn:axial_hyp_pot}.  When $\epsilon>1$ and $U_0>0$, the potential is trapping in the $zy$ plane and anti-trapping in $x$; and for $\epsilon<0$ and $U_0>0$, the potential is trapping in $z$ and $x$, but anti-trapping in $y$. Whereas, at $\epsilon=0$ and $\epsilon=1$, the potential is independent of $x$ and $y$ respectively. The frequency along the $z$-axis is
\begin{equation}
\omega_{z,\mathrm{hyp}}\equiv\sqrt{\frac{2eU_0}{ms^2}}.
\label{eqn:axial_hyp_freqz}
\end{equation}
The frequencies along the $x$ and $y$ axis are discussed in connection with the net linear microtrap potential below.

\subsection{Linear Microtrap Static Potential Analysis}

The static potential is computed using a three-dimensional finite element solver. The distance from the center of the trap to the bounding box that was used in the simulation was more then twice the tip-to-tip cantilever separation.  To reduce the error in the simulation results, several different grids were used and the results were averaged.

It is possible to approximate the three-dimensional static potential of the linear microtrap, $U_{_{\mathrm{LMT}}}(x,y,z)$ by doing a Taylor expansion about the center of the trap. Because equal voltages are applied to all capping electrodes as shown in Fig. \ref{fig:microtrap_schematic}(b), the cross-terms in the Taylor expansion are zero.  The coefficients of the harmonic terms are:
\begin{eqnarray}
D_{x}  & = & \frac{1}{U_0}\frac{\partial^{2}U_{_{\mathrm{LMT}}}}{\partial
x^{2} }(0,0,0)\\
D_{y}  & = & \frac{1}{U_0}\frac{\partial^{2}U_{_{\mathrm{LMT}}}}{\partial
y^{2} }(0,0,0)\\
D_{z}  & = & \frac{1}{U_0}\frac{\partial^{2}U_{_{\mathrm{LMT}}}}{\partial
z^{2} }(0,0,0)
\end{eqnarray}
The derivatives are then evaluated numerically on the calculated potential along the axes. The potential is therefore
\begin{eqnarray}
{U_{_{\mathrm{LMT}}}} & \approx & \frac{U_{0}}{2} \left(  D_{x} x^{2}+D_{y}y^{2}+D_{z}z^{2} \right) \nonumber\\
{}  &  =  &\frac{U_0 D_{z}}{2} \left( \frac{D_{x}}{D_{z} }x^{2} +
\frac{D_{y}}{D_{z}}y^{2}+z^{2} \right) 
\label{eqn:axial_LMT_pot1}
\end{eqnarray}
A static potential geometric efficiency factor $\kappa$ compares the static potential of the linear microtrap with the hyperbolic electrode geometry of similar characteristic dimension. The characteristic dimension of the linear microtrap that corresponds to the distance $s$ in the hyperbolic electrode geometry is the distance from the center of the trap to the nearest point on the end-cap electrodes: $d_{\mathrm{eff}}=\sqrt{\ell_{\mathrm{eff}}^2+(b/2+g)^{2}}$. 
\begin{equation}
\kappa\equiv D_{z}d_{\mathrm{eff}}^{2}/2
\end{equation}

The static potential in the linear microtrap can then be written in the same form as the potential in the hyperbolic electrode geometry.
\begin{equation}
U_{_{\mathrm{LMT}}}=\frac{U_{0}\kappa}{d_{\mathrm{eff}}^{2}}\left(  -\epsilon
x^{2}-(1-\epsilon)y^{2}+z^{2}\right)
\label{eqn:axial_LMT_pot}
\end{equation}
where $\epsilon=-D_x/D_z=1+D_y/D_z$. Given this approximation of the electrostatic potential in the linear microtrap, the form of the trap frequency along the $z$-axis is similar to that of the hyperbolic electrodes (Eq. \ref{eqn:axial_hyp_freqz}) with the difference being only the static potential geometric efficiency factor $\kappa$
\begin{equation}
\omega_{z_{,\mathrm{LMT}}}=\sqrt{\frac{{2\kappa eU_{0}}}{{md_{\mathrm{eff}}^{2}}}}.
\end{equation}

\begin{figure*}[ptbh]
\centering
\includegraphics[width=14cm,keepaspectratio]{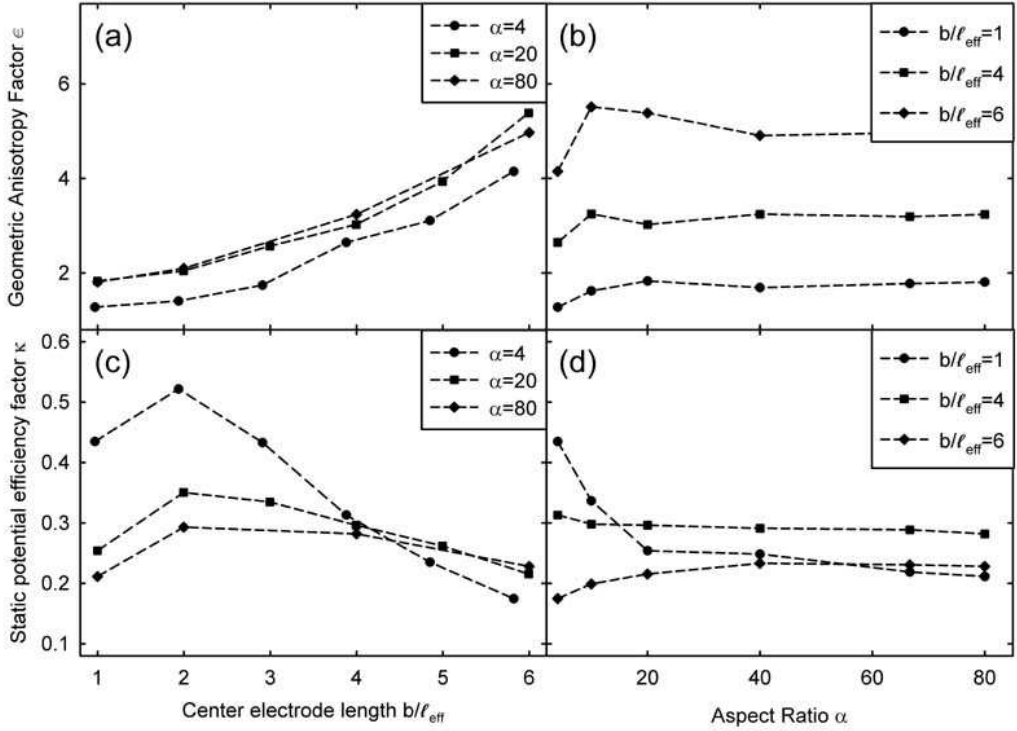}
\caption{The results for the three-dimensional numerical simulations of the static potential in the linear microtrap. Both the anisotropy factor $\epsilon$ and the static potential geometric efficiency factor $\kappa$ are shown. The ratio of the layer separation over the layer width was fixed at one and the gap separation at two (g=2 from Fig. \ref{fig:microtrap_schematic}(b)).}
\label{fig:static_data}
\end{figure*}

The results characterizing the linear microtrap for $\kappa$ and $\epsilon$ from the numerical simulations are shown in Fig. \ref{fig:static_data}.

\section{\label{sec:total_pot}Net Potential}

The combined static and ponderomotive potentials that determine the motion of a ion in the linear microtrap are written as a three-dimensional uncoupled harmonic oscillator potential:
\begin{eqnarray}
\phi_{_{\mathrm{LMT}}}&=&\psi_{_{\mathrm{LMT}}}+U_{_{\mathrm{LMT}}}\nonumber\\
{} & = &\frac{e^{2}V_{0}^{2}\eta^{2}} {4m^{2}\Omega_{T}^{2}\ell_{\mathrm{eff}}^{4}
} \left(  x^{2} + y^{2}\right)\nonumber\\
{}&{}& + \frac{\kappa U_{0}}{ d_{\mathrm{eff}}^{2} } 
\left( -\epsilon x^{2} - (1-\epsilon) y^{2}+z^{2} \right) 
\label{eqn:total_LMT_pot}
\end{eqnarray}
where the residual axial ponderomotive potential has been neglected. Considering this full potential, the effective trapping frequencies consist of the quadrature sum of the ponderomotive and the static frequencies.
\begin{eqnarray}
\omega_{x_{,\mathrm{LMT}}} & =& \sqrt{ \omega_{p_{,\mathrm{LMT}}}^2 - \epsilon\omega_{z_{,\mathrm{LMT}}}^2 } \\
& = & \sqrt{ \frac{ e^2V_0^2\eta^2 }{ 2m^2\Omega_T^2\ell_{\mathrm{eff}}^4 } - \frac{ 2\epsilon\kappa eU_0 }{ md_{\mathrm{eff}}^2 } }\nonumber\\
\omega_{y_{,\mathrm{LMT}}} & = & \sqrt{ \omega_{p_{,\mathrm{LMT}}}^2 - (1-\epsilon)\omega_{z_{,\mathrm{LMT}}}
^2 }\\
& = & \sqrt{ \frac{ e^2V_0^2\eta^2 }{ 2m^2\Omega_T^2\ell_{\mathrm{eff}}^4 }
- \frac{ 2(1-\epsilon)\kappa eU_0 }{ md_{\mathrm{eff}}^2 } }\nonumber\\
\omega_{z_{,\mathrm{LMT}}} & =& \sqrt{ \frac{ 2\kappa eU_0 }{ md_{\mathrm{eff}}^2 } }
\label{eqn:total_LMT_freq}
\end{eqnarray}
Table \ref{table:t1} provides a few examples of the calculation of the total trap frequencies given a specific geometry. The mass of the ion used in calculating the frequencies was ${}^{111}$Cd with an RF frequency of $\Omega_{T}/2\pi=50$MHz. The values for $\eta$, $\epsilon$, and $\kappa$ were taken from Figs. \ref{fig:eta_alpha} and \ref{fig:static_data}. 

\begin{center}
\begin{table*}[ptbh]
\begin{tabular}{|c|c|c|c|c|}
\hline\hline
\begin{tabular}{c}
a=40$\mu $m \\ 
d=10$\mu $m \\ 
w=10$\mu $m \\ 
b=100$\mu $m
\end{tabular}
& 
\begin{tabular}{c}
$\alpha =4$ \\ 
$\delta =1$ \\ 
$\ell _{\mathrm{eff}}=21\mu $m \\ 
$d_{\mathrm{eff}}=59\mu $m
\end{tabular}
& 
\begin{tabular}{c}
$\eta =0.7$ \\ 
$\epsilon =3$ \\ 
$\kappa =0.3$%
\end{tabular}
& 
\begin{tabular}{c}
$V_{0}=40$V \\ 
$U_{0}=20$V \\ 
$\omega_{p_{,\mathrm{LMT}}}/2\pi =20$MHz \\ 
$\omega_{z_{,\mathrm{LMT}}}/2\pi =8.7$MHz
\end{tabular}
& 
\begin{tabular}{c}
$\omega_{x_{,\mathrm{LMT}}}/2\pi =13$MHz \\ 
$\omega_{y_{,\mathrm{LMT}}}/2\pi =23$MHz \\ 
$\omega_{z_{,\mathrm{LMT}}}/2\pi =8.7$MHz
\end{tabular}
\\ \hline\hline\hline
\begin{tabular}{c}
a=40$\mu $m \\ 
d=2$\mu $m \\ 
w=2$\mu $m \\ 
b=100$\mu $m
\end{tabular}
& 
\begin{tabular}{c}
$\alpha =20$ \\ 
$\delta =1$ \\ 
$\ell _{\mathrm{eff}}=20\mu $m \\ 
$d_{\mathrm{eff}}=58\mu $m
\end{tabular}
& 
\begin{tabular}{c}
$\eta =0.43$ \\ 
$\epsilon =3.5$ \\ 
$\kappa =0.26$%
\end{tabular}
& 
\begin{tabular}{c}
$V_{0}=20$V \\ 
$U_{0}=1$V \\ 
$\omega_{p_{,\mathrm{LMT}}}/2\pi =6.7$MHz \\ 
$\omega_{z_{,\mathrm{LMT}}}/2\pi =1.8$MHz
\end{tabular}
& 
\begin{tabular}{c}
$\omega_{x_{,\mathrm{LMT}}}/2\pi =5.8$MHz \\ 
$\omega_{y_{,\mathrm{LMT}}}/2\pi =7.3$MHz \\ 
$\omega_{z_{,\mathrm{LMT}}}/2\pi =1.8$MHz
\end{tabular}
\\ \hline\hline\hline
\begin{tabular}{c}
a=80$\mu $m \\ 
d=2$\mu $m \\ 
w=2$\mu $m \\ 
b=160$\mu $m
\end{tabular}
& 
\begin{tabular}{c}
$\alpha =40$ \\ 
$\delta =1$ \\ 
$\ell _{\mathrm{eff}}=40\mu $m \\ 
$d_{\mathrm{eff}}=89\mu $m
\end{tabular}
& 
\begin{tabular}{c}
$\eta =0.38$ \\ 
$\epsilon =3.2$ \\ 
$\kappa =0.28$%
\end{tabular}
& 
\begin{tabular}{c}
$V_{0}=35$V \\ 
$U_{0}=0.9$V \\ 
$\omega_{p_{,\mathrm{LMT}}}/2\pi =2.6$MHz \\ 
$\omega_{z_{,\mathrm{LMT}}}/2\pi =1.2$MHz
\end{tabular}
& 
\begin{tabular}{c}
$\omega_{x_{,\mathrm{LMT}}}/2\pi =1.5$MHz \\ 
$\omega_{y_{,\mathrm{LMT}}}/2\pi =3.2$MHz \\ 
$\omega_{z_{,\mathrm{LMT}}}/2\pi =1.2$MHz
\end{tabular}

\\ \hline\hline
\end{tabular}
\caption{Sample calculations for trap performance.  A trap frequency of $\Omega_{T}/2\pi=50$MHz was assumed for a ${}^{111}$Cd ion.}
\label{table:t1}
\end{table*}
\end{center}

\section{\label{PA}Microtrap Principal Axes}

Principal axes are the axes along which it is possible to describe the motion of an ion in the total potential as a three-dimensional uncoupled harmonic oscillator. This means that the motion of the ion along each axis is independent of the other two spatial coordinates.  The equations of motion for an uncoupled harmonic oscillator are 
\begin{equation}
\ddot{x}=-\omega_{x}^2 x,\rm{etc}.
\end{equation}
An uncoupled harmonic oscillator corresponds to a potential with symmetries along the principal axes. Since the RF ponderomotive potential (Eq. \ref{eqn:cross_LMT_ponder}) is radially symmetric, the principal axes of a linear ion trap are determined by the static potential. The principal axes of an ion trap are of concern when considering laser cooling an ion in the trap. Laser cooling along all three dimensions of motion is possible only if the laser wave vector $\vec{k}_{laser}$ has a vector component along all three principal axes. The symmetry of the microtrap is such that the $z$-axis is a principal axis, therefore, the axes of concern are in the $xy$ plane. It is possible to rotate the principal axes by applying different static voltages to the electrodes, which give rise to an $xy$ cross-term in the static potential.

To find the new principal axes, one can rotate the coordinate system via Eq. \ref{eqn:rot_theta}.
\begin{eqnarray}
x  & =&x^{\prime}\cos\theta+y^{\prime}\sin\theta\nonumber\\
y  & =&-x^{\prime}\sin\theta+y^{\prime}\cos\theta
\label{eqn:rot_theta}%
\end{eqnarray}
This rotation can be applied to the potential with an $xy$ cross-term of magnitude $\lambda$
\begin{equation}
U^{\prime}_{_{\mathrm{LMT}}}=\frac{\kappa^{2}U_{0}}{d_{\mathrm{eff}}^{2}}\left(
-\epsilon x^{2}-(1-\epsilon)y^{2}+\lambda xy+z^{2}\right)
\end{equation}
to find an angle at which the cross term in the rotated coordinate system ($\lambda^{\prime} x^{\prime}y^{\prime}$) vanishes.  This new coordinate system, rotated about the $z$-axis by an angle $\theta$, now determines the principal axes of the trap. The angle at which the cross-term vanishes is found as a function of the coefficient of the cross-term $\lambda$ and the geometric factor $\epsilon$.
\begin{equation}
\tan(2\theta)=\frac{\lambda}{2\epsilon-1}
\label{eqn:theta_cond}
\end{equation}
The static axial potential can then be written in an uncoupled form, showing explicitly the new principal axes $x^{\prime}$ and $y^{\prime}$.
\begin{equation}
U^{\prime}_{_{\mathrm{LMT}}}=\frac{\kappa^{2}U_{0}}{d_{\mathrm{eff}}^{2}}\left(
-\epsilon^{\prime}x^{\prime}{}^{2}-(1-\epsilon^{\prime}) y^{\prime}{}
^{2}+z^{2}\right)
\end{equation}
where $\epsilon^{\prime}=\epsilon\cos(2\theta)+\frac{\lambda}{2}\sin(2\theta)+\sin^{2}\theta$.

\begin{figure}[ptbh]
\centering
\includegraphics[width=7cm,keepaspectratio]{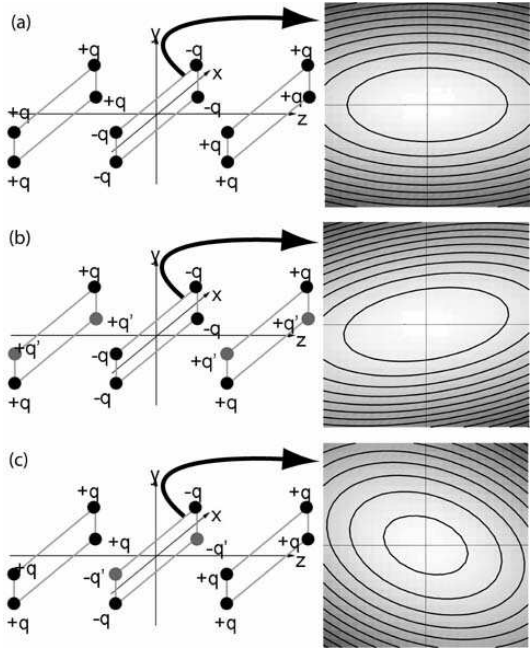}
\caption{(a)Unrotated static 12 point charge potential, shown as cross section in the $z=0$ plane. (b) By changing the charge on four of the eight end-cap points, the principal axes rotate. (c) The same result can be achieved by applying additional negative charge to the center electrodes. The aspect ratio of $\alpha=4/3$ was used to illustrate the rotation of the axes of symmetry of the potential.}
\label{fig:rotate}
\end{figure}

A simple point charge potential model can be used to provide a qualitative idea of how the principal axes may be rotated.  Twelve charges are fixed at the corners of three rectangles as shown in Fig. \ref{fig:rotate}(a).  The positions of eight charges of value $+q$ are at $(\pm a/2,\pm d/2, \pm b)$ and an additional four with charge $-q$ at $(\pm a/2,\pm d/2, 0)$. A Taylor expansion of the point charge potential where $b\gg a,d$ can be written as 
\begin{equation}
U_{\mathrm{point}}=\frac{U_{0}}{r_{0}^{2}}\left(  -\epsilon x^{2} -(1-\epsilon) y^{2} + z^{2} \right)
\end{equation}
where $U_0=2q/(4\pi\varepsilon_0r_0)$, $r_0=\sqrt{(a/2)^2+(d/2)^2}$, and $\epsilon=(2a^2-d^2)/(a^2+d^2)$.  If two charges are increased from $q$ to  $q^{\prime}$ on either end-cap as in Fig. \ref{fig:rotate}(b), the principal axes are rotated.  Alternatively, one could increase the negative charge on two of the four point charges in the $z=0$ plane.  This would correspond to applying a negative static potential to two of the center electrodes in the linear microtrap and is more effective at rotating the principal axes.  The potential in the point charge model, with the addition of these modified charges, becomes
\begin{equation}
U_{\mathrm{point}}=\frac{2(q+q^{\prime})}{(4\pi\varepsilon_{0})r_{0}^{3}}\left(  -\epsilon x^{2}-(1-\epsilon)y^{2}+\lambda xy + z^{2}\right) 
\label{eqn:point_pot_2}
\end{equation}
where now, the $xy$ cross term has a coefficient
\begin{equation}
\lambda=6\frac{q-q^{\prime}}{q+q^{\prime}}\frac{ad}{a^2+d^2}
\label{eqn:lambda}
\end{equation}
Substituting Eq. \ref{eqn:lambda} into the condition for the rotation angle (Eq. \ref{eqn:theta_cond}), and using the explicit form for $\epsilon$ in the point charge model, the rotation angle can be expressed as a function of the applied charges and the trap aspect ratio ($\alpha=a/d$, the tip-to-tip cantilever separation over the layer spacing).
\begin{equation}
\tan2\theta=\frac{1-q^{\prime}/q}{1+q^{\prime}/q}\frac{2\alpha}{\alpha^2-1}
\label{eqn:center_rot_scheme}
\end{equation} 
There are several features of this model that give a qualitative understanding of the rotation of the principal axes.  First, for a given trap aspect ratio $\alpha$, by increasing the ratio of charges, one can rotate the principal axes a fixed amount.  However, as the aspect ratio increases, the amount of rotation that can be given the principal axes by changing the charge ratio is decreased, eventually approaching zero.

\begin{figure}[ptbh]
\centering
\includegraphics[width=7cm,keepaspectratio]{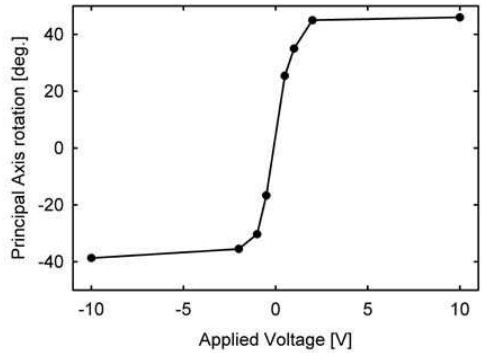}
\caption{The principal axis rotation from the $xy$ axis shown in Fig. \ref{fig:microtrap_schematic}(a) as a function of the applied voltage on two diagonally opposing center electrodes.  The other center electrodes were held at static ground with the end-cap voltages fixed at $U_0=1$V.  The trap dimensions are $a=40\mu$m, $d=2\mu$m, w=$2\mu$m, with all electrodes having a width of $100\mu$m.}
\label{fig:lmt_axis_rot}
\end{figure}

The principal axes rotation in the $xy$ plane for the linear microtrap as a function of the applied voltage on two diagonally opposing center electrodes is shown in Fig. \ref{fig:lmt_axis_rot}. The aspect ratio was fixed at $\alpha=20$ and $\delta=1$. The other two center electrodes were held at static ground with all eight end-cap electrodes at $U_0=1$V. As discussed above, by applying small voltages to the appropriate center electrodes, it is possible to rotate the principal axes so that laser cooling is effective.

\section{Conclusion}

A new design for a microfabricated linear ion trap has been discussed. Calculations of the RF ponderomotive potential have shown a surprising degree of isotropy near the center of the trap, even for very high aspect ratios.  We find that for high transverse electrode aspect ratios, the trap strength approaches $1/\pi$ times that of a comparable hyperbolic electrode structure. This may be of importance in the design of microtraps in applications such as Cavity QED \cite{berman} and miniature mass spectrometers where conventional ion trap designs can not be used. Geometric scaling factors for the linear microtrap provide an easy comparison between these new trap designs and conventional ion traps, facilitating implementation in future experiments.

Further investigations will require actual fabrication and experimentation with this new type of trap and include an investigation of the patch potentials on the surfaces of the doped semiconductors, the limiting electric field, and laser scatter from the small aperture.  These factors are all technical in nature and should not prohibit the future implementation of this novel linear microtrap design.

\appendix
\section{\label{sec:analytic}Appendix: Analytic Solution of the Transverse Potential}

\begin{figure}[ptbh]
\centering
\includegraphics[width=7cm,keepaspectratio]{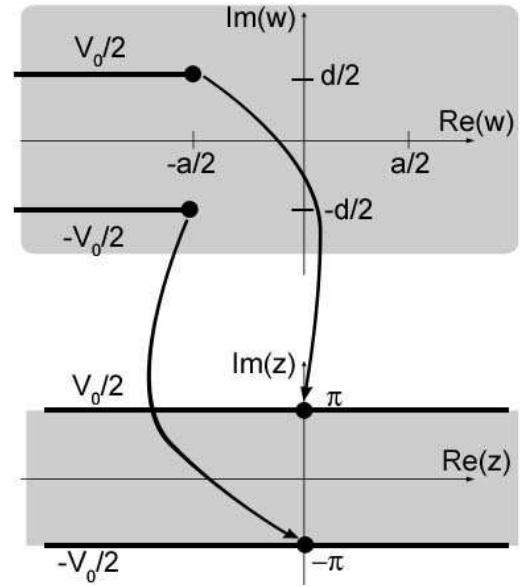}\caption{The linear microtrap model in the complex $w$ plane with semi-infinite electrodes that terminate at $\pm a/2\pm id/2$ with applied voltages $\pm V_{0}/2$.}
\label{fig:analytic}
\end{figure}
The cross-section of the linear microtrap can also be modeled as semi-infinite electrodes in a complex plane.  This model enables calculation of an analytic solution for the geometric factor $\eta$ in the limit of infinitely thin electrodes. Following the analysis of parallel-plate capacitor fringe fields of Valluri et.al \cite{valluri:2000}, the cross-section of the left cantilever electrodes are described in the complex plane as lines that go from negative infinity along the real axis and terminate at $-a/2\pm id/2$, as shown in Fig. \ref{fig:analytic}. The right set of electrodes (not shown) are a mirror image across the $x=0$ line and terminate at $a/2\pm id/2$. The electrodes are then mapped to an infinite parallel plate capacitor.  The function that does this mapping is
\begin{equation}
\pm\frac{2w\pi}{d}+\frac{a\pi}{d}-1=z+e^{z}.
\label{eqn:mapz_w}
\end{equation}
The positive value maps the parallel plate capacitor to the left set of electrodes in the $w$ plane, and the negative corresponds to the right set. The potential in the strip between the two electrodes in the $z$ plane is simply the potential between two parallel plates in a capacitor, written in complex notation:
\begin{equation}
\Phi=\frac{V_0}{2\pi}\mathrm{Im}(z),
\end{equation}
where $\mathrm{Im}(z)$ denotes the imaginary part of $z$.

To find the potential of the original electrode geometry, the inverse function of Eq. \ref{eqn:mapz_w} is needed.  With that inverse map the potential in the $w$ plane can be evaluated. The inverse map can be written in terms of the Lambert W function, $\mathrm{W}_{k}(\xi)$, following \cite{valluri:2000}.
\begin{equation}
z_{\pm}=\zeta_{\pm}-\mathrm{W}_{k}\left(  e^{\zeta_{\pm}}\right)
\label{eqn:mapw_z}
\end{equation}
where $\zeta_{\pm}=\pm\frac{2w\pi}{d}+\frac{a\pi}{d}-1$ is a scaled complex variable. The Lambert W function $y=\mathrm{W}_{k}(x)$ is the solution to the equation
$x=y\exp{y}$.  For complex variables, it is important to select the proper branch of $\mathrm{W}_{k}(\xi)$ when evaluating the function. The appropriate branch is found using \cite{valluri:2000} 
\begin{equation}
k=\left\lceil \frac{\mathrm{Im} (\zeta)-\pi}{2\pi}\right\rceil,
\end{equation}
where $\left\lceil \right\rceil $ denotes the ceiling function which indicates that the argument inside the ceiling function should be rounded up to the nearest integer.

If the tip-to-tip cantilever separation $a$ is much greater then the layer separation $d$ ($\alpha=a/d\gg1$), the potential at the center of the trap can be approximated as the linear combination of the potential from both the left electrodes and the right electrodes
\begin{equation}
\Phi=\frac{V_0}{2\pi} \left(\mathrm{Im}(z_{+}) + \mathrm{Im}(z_{-})\right).
\label{eqn:phi_anl}
\end{equation}
With this approximation ($\alpha\gg1$) an asymptotic form of the Lambert W function exists that leads to a simplification of the inverse map Eq. \ref{eqn:mapw_z}. The principal branch of the Lambert W function has an asymptotic form:
\begin{equation}
\mathrm{W}_{0}(\xi)\approx\ln\xi-\ln(\ln\xi),\xi\gg1.
\label{eqn:w_asympt}
\end{equation}
Inserting Eq. \ref{eqn:w_asympt} into Eq. \ref{eqn:mapw_z}, the inverse map becomes: $z\approx\ln\zeta_{\pm}$. Expanding the log function about $w=0$, Eq. \ref{eqn:mapw_z} can be written:
\begin{equation}
z_{\pm}=\ln(\frac{a\pi}{d}-1)+\left[  \pm\frac{2\pi w}{a\pi-d}-\frac{1}
{2}\left[  \frac{2\pi w}{a\pi-d}\right]  ^{2}+\dots\right].
\label{eqn:z_expand}
\end{equation}
Since the potential is the linear combination of $\mathrm{Im}(z_{+})$ and $\mathrm{Im}(z_{-})$ (Eq. \ref{eqn:phi_anl}) and the linear terms are opposite in sign, only the quadratic term contributes to the potential of the microtrap. Squaring the complex variable $w=u+iv$ and keeping only the second-order imaginary terms, one finds that the potential is 
\begin{equation}
\Phi=-\frac{4\pi V_0}{\left(  a\pi-d\right)  ^{2}}uv.
\label{eqn:analytic_pot_unrot}
\end{equation}
By rotating the coordinate system about the origin by $\theta=\pi/4$, the potential is written in a form that allow for easy comparison with the quadrupole potential of Eq. \ref{eqn:cross_hyp_pot}:
\begin{equation}
\Phi=\frac{2\pi V_0}{\left(  a\pi-d\right)  ^{2}}\left(  u'{}^{2}-v'{}^{2}\right)  .
\label{eqn:analytic_pot_rot}
\end{equation}
The geometric factor $\eta$ can be found for a microtrap with effective distance $\ell_{\mathrm{eff}}=\sqrt{(a/2)^{2}+(d/2)^{2}}$. 
\begin{eqnarray}
\eta&=&\frac{4\pi}{\left(  a\pi-d\right)  ^{2}}\ell_{\mathrm{eff}}^{2}\nonumber\\
&=&\pi \frac{\alpha^{2}+1}{\left(  \alpha\pi-1\right)  ^{2}}.
\label{eqn:analytic_eta}
\end{eqnarray}
This analytic solution of the geometric factor is valid in the limit where the trap aspect ratio is large: when the tip-to-tip cantilever separation is much larger then the layer separation.  The geometric factor asymptotically approaches $\eta=1/\pi$ in this limit.  The analytic solution (Eq. \ref{eqn:analytic_eta}) is shown as the solid line in Fig. \ref{fig:eta_alpha}.  Note that this complex model assumes infinitely thin electrodes which correspond to a large value for the ratio of the layer separation to the layer thickness $\delta=d/w\rightarrow\infty$.  The values for $\eta$ found via numerical simulations approach the analytic solution as $\delta$ increases and also approach the asymptotic value of $\eta=1/\pi$ for large $\alpha$.

In addition, the analytic model can be used to calculate the asymptotic values for the ponderomotive potential depth and the maximum trap size along the weak axis $r_{max}$.  Inserting the asymptotic form of the Lambert W function (Eq. \ref{eqn:w_asympt}) directly into the potential (Eq. \ref{eqn:phi_anl}) and evaluating the imaginary part, the potential can be written directly as a function of $u$ and $v$.
\begin{eqnarray}
\Phi&=&\frac{V_0}{2\pi}\left[\mathrm{tan}^{-1}\left({\frac{2v}{2u+a-d/\pi}}\right)\right.\nonumber\\
{}&{}&\left.+\mathrm{tan}^{-1}\left({\frac{-2v}{-2u+a-d/\pi}}\right)\right]
\label{eqn:phi_anl_approx}
\end{eqnarray}
The pseudopotential can then be directly evaluated, using a two-dimensional gradient, from Eq. \ref{eqn:ponder_defn}.  The maximum of the pseudopotential along the $v$-axis ($v=r_{\mathrm{max}}$) lies at
\begin{eqnarray}
r_{\mathrm{max}}&=&\frac{1}{2\pi}(a\pi-d)\nonumber\\
&=&\frac{a}{2}(1-\frac{1}{\pi\alpha}).
\label{eqn:v_max}
\end{eqnarray}
The location of the potential maximum asymptotically approaches $r_{\mathrm{max}}=a/2$ as the aspect ratio goes to infinity. The trap depth is the pseudopotential evaluated at this maximum:
\begin{equation}
\psi(r_{\mathrm{max}})=\frac{e^2V_0^2}{4m\Omega_{T}^{2}}\frac{1}{a^2\pi^2\left(1-\frac{1}{\alpha\pi}\right)^2}.
\label{eqn:analytic_depth}
\end{equation}
The analytic solution for the scaled trap depth is shown in Fig. \ref{fig:trap_depth} and approaches the asymptotic value of $2694$ [K$\cdot\mu$m$^2$/V$^2$] for large $\alpha$.

\begin{acknowledgement}
We are pleased to acknowledge useful discussions with Keith Schwab.  This work is supported by the Advanced Research and Development Activity and the National Security Agency under Army Research Office contract DAAD 19-01-1-0667, the National Science Foundation ITR Program, and the NIST Small Business Innovative Research Program.
\end{acknowledgement}

\bibliographystyle{unsrt}
\bibliography{planar_ion_trap_paper_bib}

\begin{thebibliography}{10}

\bibitem{paul:1990}
W.~Paul.
\newblock Electromagnetic traps for charged and neutral particles.
\newblock {\em Rev. Mod. Phys}, 62:531--540, 1990.

\bibitem{fisk:1997}
P.~T.~H. Fisk.
\newblock Trapped-ion and trapped-atom microwave frequency standards.
\newblock {\em Reports on Progress in Physics,}, 60:761--817, 1997.

\bibitem{vandyck:1987}
{R.S. Van Dyck Jr.}, {P.B. Pschwinberg}, and {H.G. Dehmelt}.
\newblock New high-precision comparison of electron and positron g factors.
\newblock {\em Phys. Rev. Lett.}, 59:26--29, 1987.

\bibitem{leibfried:2003}
D.~Leibfried, R.~Blatt, C.~Monroe, and D.~Wineland.
\newblock Quantum dynamics of single trapped ions.
\newblock {\em Rev. Mod. Phys}, 75:281--324, 2003.

\bibitem{steane:1997}
A.~Steane.
\newblock The ion trap quantum information processor.
\newblock {\em Appl. Phys. B}, 64:623--642, 1997.

\bibitem{wineland:1998}
D.~J. Wineland, C.~Monroe, W.~M. Itano, D.~Leibfried, B.~E. King, and D.~M.
  Meekhof.
\newblock {Experimental issues in coherent quantum-state manipulation of
  trapped atomic ions}.
\newblock {\em Journal of Research of the National Institute of Standards and
  Technology}, 103:259--328, 1998.

\bibitem{kielpinski:2002}
D.~Kielpinski, C.~Monroe, and D.J. Wineland.
\newblock Architecture for a large-scale ion-trap quantum computer.
\newblock {\em Nature}, 417:709--711, 2002.

\bibitem{ye:1999}
J.~Ye, D.~W. Vernooy, and H.~J. Kimble.
\newblock {Trapping of Single Atoms in Cavity QED}.
\newblock {\em Phys. Rev. Lett.}, 83:4987--4990, 1999.

\bibitem{pinkse:2000}
P.~W.~H. Pinkse, T.~Fischer, P.~Maunz, and G.~Rempe.
\newblock Trapping an atom with single photons.
\newblock {\em Nature}, 404:365--368, 2000.

\bibitem{guthoehrlein:2002}
G.~R. Guth{\"{o}}hrlein, M.~Keller, K.~Hayasaka, W.~Lange, and H.~Walther.
\newblock A single ion as a nanoscopic probe of an optical field.
\newblock {\em Nature}, 414:49, 2001.

\bibitem{mundt:2002}
A.B. Mundt, A.~Kreuter, C.~Becher, D.~Leibfried, J.~Eschner, F.~Schmidt-Kaler,
  and R.~Blatt.
\newblock Coupling a single atomic quantum bit to a high finesse optical
  cavity.
\newblock {\em Phys. Rev. Lett.}, 89:103001, 2002.

\bibitem{taylor:2001}
S.~Taylor, R.~F. Tidnall, and R.~R.~A. Syms.
\newblock Silicon based quadrupole mass spectrometry using
  microelectromechanical systems.
\newblock {\em J. Vac. Sci. Tech.}, 19:557, 2001.

\bibitem{folman:2002}
R.~Folman, P.~Kr{\"{u}}ger, J.~Schmiedmayer, J.~Denschlag, and C.~Henkel.
\newblock Microscopic atom optics: From wires to an atom chip.
\newblock {\em Advances in Atomic, Molecular and Optical Physics}, 48:263--356,
  2002.

\bibitem{henkel:1999}
C.~Henkel, S.~P{\"{o}}tting, and M.~Wilkens.
\newblock Loss and heating of particles in small and noisy traps.
\newblock {\em Appl. Phys. B}, 69:379--387, 1999.

\bibitem{turchette:2000}
Q.~A. Turchette, D.~Kielpinski, B.~E. King, D.~Leibfried, D.~M. Meekhof, C.~J.
  Myatt, M.~A. Rowe, C.~A. Sackett, C.~S. Wood, W.~M. Itano, C.~Monroe, and
  D.~J. Wineland.
\newblock Heating of trapped ions from the quantum ground state.
\newblock {\em Phys. Rev. A}, 61:063418, 2000.

\bibitem{witteborn:1977}
F.~C. Witteborn and W.~M. Fairbank.
\newblock Apparatus for measuring the force of gravity on freely falling
  electrons.
\newblock {\em Rev. Sci. Instr.}, 48:1--11, 1977.

\bibitem{raizen:1992}
M.~G. Raizen, J.~M. Gilligan, J.~C. Bergquist, W.~M. Itano, and D.~J. Wineland.
\newblock Ionic crystals in a linear paul trap.
\newblock {\em Phys. Rev. A}, 45:6493, 1992.

\bibitem{cleveland:1993}
J.~P. Cleveland, S.~Manne, D.~Bocek, and P.~K. Hansma.
\newblock A nondestructive method for determining the spring constant of
  cantilevers for scanning force microscopy.
\newblock {\em Rev. Sci. Instr.}, 64:403, 1993.

\bibitem{lifshitz:2000}
R.~Lifshitz and M.~L. Roukes.
\newblock Thermoelastic damping in micro- and nanomechanical systems.
\newblock {\em Phys. Rev. B}, 61:5600--5609, 2000.

\bibitem{david:1996}
J.P.R. David and G.E. Stillman.
\newblock {\em {Carrier Ionisation Coefficients of GaAs}}, chapter 4.9, pages
  190--197.
\newblock INSPEC, The Institution of Electrical Engineers., 1996.

\bibitem{rauthan:1992}
C.~M.~S. Rauthan and J.~K. Srivastava.
\newblock Electrical breakdown voltage characteristics of buried silicon
  nitride layers and their correlation to defects in the nitride layer.
\newblock {\em Materials Letters}, 9:252, 1990.

\bibitem{dehmelt:1967}
H.G. Dehmelt.
\newblock Radiofrequency spectroscopy of stored ions i: Storage.
\newblock {\em Adv. At. Mol. Phys.}, 3:53, 1967.

\bibitem{syms:1998}
R.~R.~A. Syms, T.~J. Tate, M.~M. Ahmad, and S.~Taylor.
\newblock {Design of a Microengineered Electrostatic Quadupole Lens}.
\newblock {\em IEEE Transactions on Electron Devices}, 45:2304, 1998.

\bibitem{berman}
P.~Berman, editor.
\newblock {\em Cavity QED}.
\newblock Adv. At. Molec. Opt. Phys. Supplement 2. Academic Press, 1994.

\bibitem{valluri:2000}
S.~R. Valluri, D.~J. Jeffery, and R.~M. Corless.
\newblock {Some Applications of the Lambert W Function to Physics}.
\newblock {\em Can. J. Phys.}, 78:823--831, 2000.

\end{thebibliography}
\end{document}